\documentclass[journal=nalefd,manuscript=letter]{achemso}
\usepackage[version=3]{mhchem}
\usepackage{amsmath}
\usepackage{amsfonts}
\usepackage{amssymb}
\usepackage{graphicx}
\usepackage{color}

\title[] {Using ultra-thin parylene films as an organic gate insulator in nanowire field-effect transistors}

\author{J.G. Gluschke}
\affiliation{School of Physics, University of New South Wales, Sydney NSW 2052, Australia}

\author{J. Seidl}
\affiliation{School of Physics, University of New South Wales, Sydney NSW 2052, Australia}

\author{R.W. Lyttleton}
\affiliation{School of Physics, University of New South Wales, Sydney NSW 2052, Australia}

\author{D.J. Carrad}
\affiliation{School of Physics, University of New South Wales, Sydney NSW 2052, Australia}
\alsoaffiliation{Center for Quantum Devices, Niels Bohr Institute, University of Copenhagen, Copenhagen DK-2100, Denmark}

\author{J.W. Cochrane}
\affiliation{School of Physics, University of New South Wales, Sydney NSW 2052, Australia}

\author{S. Lehmann}
\affiliation{Solid State Physics/NanoLund, Lund University SE-221 00 Lund, Sweden}

\author{L. Samuelson}
\affiliation{Solid State Physics/NanoLund, Lund University SE-221 00 Lund, Sweden}

\author{A.P. Micolich}
\email{adam.micolich@nanoelectronics.physics.unsw.edu.au}
\affiliation{School of Physics, University of New South Wales, Sydney NSW 2052, Australia}

\date{\today}

\begin{document}

\begin{abstract}

We report the development of nanowire field-effect transistors featuring an ultra-thin parylene film as a polymer gate insulator. The room temperature, gas-phase deposition of parylene is an attractive alternative to oxide insulators prepared at high temperatures using atomic layer deposition. We discuss our custom-built parylene deposition system, which is designed for reliable and controlled deposition of $< 100$~nm thick parylene films on III-V nanowires standing vertically on a growth substrate or horizontally on a device substrate. The former case gives conformally-coated nanowires, which we used to produce functional $\Omega$-gate and gate-all-around structures. These give sub-threshold swings as low as $140$~mV/dec and on/off ratios exceeding $10^3$ at room temperature. For the gate-all-around structure, we developed a novel fabrication strategy that overcomes some of the limitations with previous lateral wrap-gate nanowire transistors. Finally, we show that parylene can be deposited over chemically-treated nanowire surfaces; a feature generally not possible with oxides produced by atomic layer deposition due to the surface `self-cleaning' effect. Our results highlight the potential for parylene as an alternative ultra-thin insulator in nanoscale electronic devices more broadly, with potential applications extending into nanobioelectronics due to parylene's well-established biocompatible properties.

{\bf Keywords:} parylene, nanowires, gate-all-around, field-effect transistor, organic gate insulator
\end{abstract}

\maketitle

Thin oxide films are near ubiquitous as the gate insulator layer in nanoscale transistors. These oxides, e.g., \ce{SiO2}, \ce{Al2O3} and \ce{HfO2} are typically deposited by atomic layer deposition (ALD) or plasma-enhanced chemical vapor deposition (PECVD) to give thin pinhole-free coatings with high dielectric constant.\cite{WongMicroelecEng06, RobertsonMSER15, RielMRS14} However oxide insulators have two major drawbacks in devices such as III-V nanowire transistors. Firstly, the high density of charge traps at the semiconductor-oxide interface causes gate hysteresis,\cite{DayehAPL07, RoddaroAPL08, HollowayJAP13} drift and instability,\cite{HollowayJAP13} reduced carrier mobility,\cite{LiuAPL11} reduced gate capacitance\cite{DayehJVSTB07} and noise.\cite{HollowayJAP13} These oxide-related problems extend to semiconductor materials beyond III-Vs, e.g., organic semiconductors,\cite{FleischliLangmuir10} 2D materials,\cite{Illarionov2DM17, BonmannJVSTB17} etc. Secondly, oxide deposition involves chemisorption and covalent bonding to the semiconductor surface. Many pristine and chemically treated surfaces are destroyed or severely damaged\cite{MilojevicAPL08, HinkleAPL08} during deposition or the oxide cannot be deposited at all.\cite{FarmSST12, DongSciRep14, AvilaAMI14} This provides a strong incentive to explore alternative materials for use as nanoscale gate insulators. In this paper, we demonstrate the potential of the organic polymer parylene (poly-p-xylylene) as a gate insulator for nanoscale devices, using III-V nanowire transistors as a test platform.

Parylene has properties that are desirable for use as a gate insulator in nanoscale transistors. First and foremost parylene can be deposited from the gas phase to give a conformal coating, unlike many other polymer insulators, which can only be deposited from solution. This means it can be deployed as a direct replacement for the ALD process most commonly used to deposit oxides for nanoscale transistors. Secondly, parylene deposition occurs by physisorption\cite{FortinCM02} rather than the chemisorption that occurs with oxides. This enables parylene deposition over a chemically passivated surface.\cite{VaethLangmuir00}  Thirdly, parylene is biocompatible\cite{ChangLangmuir07} and widely used in biomedical applications, e.g., neural interfacing,\cite{KhodagholyNatComm13, KoitmaeAMI16} offering new opportunities for, e.g., nanowire-based nanobioelectronics.\cite{NoyAdvMat10, ZhangChemRev16} An additional advantage of parylene is that the sample remains at room temperature throughout deposition avoiding thermal damage or annealing effects in thermally-sensitive materials, e.g., organic semiconductors. For example, Podzorov {\it et al.} used a $200$~nm thick parylene film to make transistors using notoriously fragile organic molecular crystals such as rubrene to obtain improved carrier mobility through reduced charge-trap density.\cite{PodzorovAPL03,PodzorovAPL03b} Parylene has also been used as a intermediate coating between the channel and oxide insulator in graphene\cite{SabriAPL09} and \ce{MoSe2}\cite{ChamlagainACSNano14} transistors in attempts to overcome undesirable aspects of direct deposition on the bare \ce{SiO2} surface in these devices.

The use of parylene in nanoscale devices presents some challenges; the key one being film thickness. The gate insulator in a nanoscale device is typically much less than $100$~nm thick and as thin as $12$~nm for the conformal insulator in a nanowire wrap-gate transistor.\cite{StormNL12} However, most commercially available parylene coating systems are designed to give films with thicknesses from $200$~nm to many microns with a focus on uniformity over large coating area. Ultra-thin ($<100$~nm) parylene films are typically obtained by limiting the amount of dimer precursor loaded into the sublimation chamber to milligram quantities.\cite{RappTSF12} This can make the control required for nanoscale device applications difficult and capricious.

In this paper we report the development of a gate-all-around InAs nanowire transistor featuring a conformal sub-$30$~nm thickness parylene gate insulator. Parylene was deposited on both free-standing nanowires and nanowires already transferred to a substrate, using a custom-built parylene deposition system designed to consistently give sub-$100$~nm films. The deposition system was designed with several features to provide nm-level thickness control irrespective of starting dimer load. We present a reliable process for device fabrication that gives excellent electrical performance, e.g., sub-threshold slopes as low as $140$~mV/dec, with high yield. The process we develop is widely applicable to nanoscale devices featuring other nanomaterials. Finally, we demonstrate that parylene deposition is compatible with retaining simple ALD-incompatible surface chemistry on an InAs surface and that a parylene coating alone can significantly reduce gate hysteresis in InAs nanowire transistors under ambient conditions. Our results highlight the strong potential for using parylene in nanoscale devices as both a conformal gate insulator and biocompatible coating to advance applications of nanowire transistors and other nanoscale devices in nanobioelectronics.

\begin{figure}
\includegraphics[width=8.5cm]{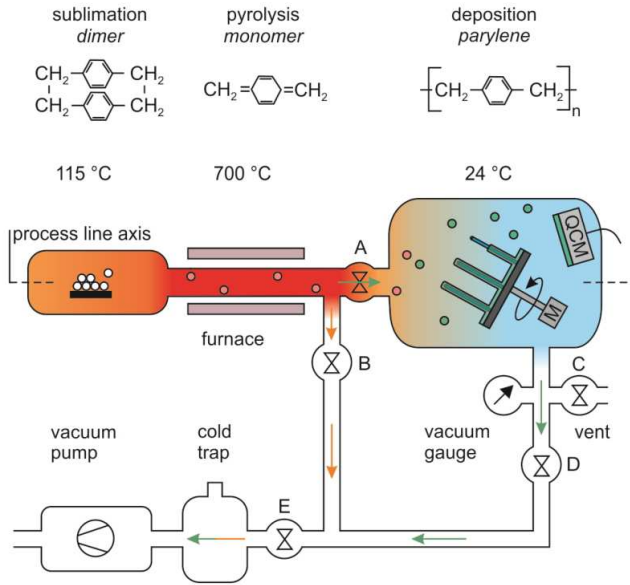}
\caption{Schematic illustrating the basic design and operation of our parylene deposition system. A precursor dimer is sublimated and drifts by pressure gradient into the furnace where it is cracked into reactive monomers. The monomers physisorb onto the cooler surfaces in the deposition chamber and polymerize. The sample is mounted on a rotating stage driven by an electric motor (M) and oriented at $30^{^{\circ}}$ to the process line axis to ensure an even conformal coating. Deposition is monitored with a quartz crystal microbalance (QCM) and is rapidly terminated by closing Valve A and opening Valve B to divert the monomer flow and evacuate the deposition chamber.}
\end{figure}

Figure~1 shows a simple schematic of our custom-built parylene system. Deposition proceeds by sublimation of the precursor dimer di-para-xylylene at $115^{\circ}$C, pyrolysis into reactive monomers at $\sim700^{\circ}$C, and room temperature deposition/polymerization at the sample under moderate vacuum ($\sim10^{-2}$~mbar).\cite{FortinBook03} The key features for enacting nm-level thickness control are: a quartz crystal microbalance (QCM) sensor in the deposition chamber for real-time thickness monitoring,\cite{RappTSF12} a small deposition chamber equipped with a valve to the process line, and a parallel `diversion' path around the deposition chamber. Our rationale was to rapidly terminate deposition once the desired thickness was achieved rather than rely on accurate control over precursor mass, as is commonly used.\cite{RappTSF12} One approach is to simply terminate dimer sublimation but this can be slow, and combined with already sublimated dimer, result in significant overshoot error. Instead we divert the normal parylene process flow (green arrows in Fig.~1) around the deposition chamber by closing Valve A and opening Valve B to terminate deposition. The alternate pathway (orange arrows in Fig.~1) enables us to terminate sublimation and deposition at a sensible timescale without risk of adverse pressure building up inside the system. The small deposition chamber volume ($\sim5\times10^{-4}$~m$^{3}$) combined with pumping via Valve D helps to ensure rapid process termination. Samples were mounted on a motor-driven rotating stage oriented at a $30^{\circ}$ angle to the process line axis to increase coating uniformity. The deposition pressure was $\sim1\times10^{-2}$~mbar giving a typical deposition rate of $\sim0.5$~\AA/s. A more detailed discussion of the design, construction and performance of our parylene deposition system will be published elsewhere.\cite{Gluschke18}

\begin{figure}
\includegraphics[width=8.5cm]{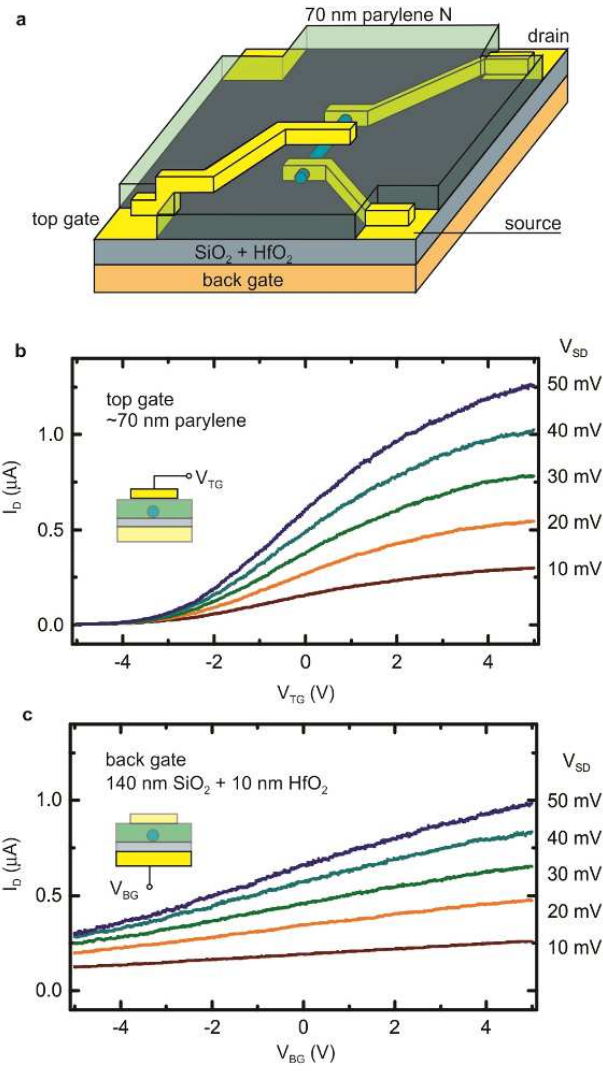}
\caption{(a) Schematic of an InAs nanowire field-effect transistor with a parylene-insulated top-gate. (b) Drain current $I_{D}$ vs top-gate voltage $V_{TG}$ at different source-drain biases $V_{SD}$. (c) $I_{D}$ vs back-gate voltage $V_{BG}$ at different $V_{SD}$ for the same device. The inactive gate was grounded in each case. All data obtained at room temperature.}
\end{figure}

We began by making a simple top-gated nanowire field-effect transistor (NWFET) to test the feasibility of using ultra-thin parylene films in nanowire devices (Fig.~2(a)). The device was fabricated on a $n^{+}$-doped Si substrate with a $140/10$~nm thick \ce{SiO2}/\ce{HfO2} layer. The $n^{+}$-Si substrate was used as a back-gate to confirm gate operation and provide a preliminary comparative assessment of parylene dielectric performance relative to the more established oxides. Fabrication began with producing a traditional back-gated NWFET\cite{DuanNat01} and then coating with $\sim70$~nm of Parylene-N (See Supplementary Figure~1). The parylene needed to be removed in select areas to enable electrical contact to the top-gate. Removal was performed by oxygen plasma etching,\cite{MengJMM08} which is necessary due to the excellent solvent resistance of parylene.\cite{FortinBook03} The solvent resistance of parylene is actually a major advantage as it enables conventional UV and electron-beam lithography processes -- including metal lift-off -- to be performed without damaging the parylene. After exposing the contact pads, the top-gate was produced using electron-beam lithography (EBL), thermal evaporation of Ti/Au and lift-off in acetone.

We compare operation of the NWFET under top-gate and back-gate control in Figs.~2(b) and 2(c), respectively. Unless otherwise indicated, the sweep gate voltage is swept from positive to negative values. The device gives a drain current $I_{D}\sim1~\mu$A at source-drain bias $V_{SD} = 50$~mV when the voltage applied to the top-gate $V_{TG}$ and back-gate $V_{BG}$ are both zero. Gating is observed in both cases (Fig.~2(b/c)) with better performance obtained for the parylene top-gate. This is expected given the comparative dielectric thickness ($70$~nm for parylene versus $150$~nm for the oxide bilayer), which dominates over the slight disparity in dielectric constant ($\epsilon_{p}~=~2.7$ for parylene versus $\epsilon_{ox}~=~4.1$ for the oxide\cite{StormNL12}). The top-gate in this device gives sub-threshold swing $S = 1.5$~V/dec and a threshold voltage $V_{th} = -2.6$~V at $V_{SD} = 50$~mV. We find low gate leakage current $I_{TG} < 50$~pA is obtained with high yield, indicating our parylene films are pinhole-free. This is consistent with previous work by Spivack \& Ferrante\cite{SpivackJECS69} and Rapp {\it et al.}\cite{RappTSF12} showing that parylene forms closed films for thicknesses exceeding $35$~nm. An important advantage regarding pinholes for nanoscale devices with even thinner parylene films is that the active interfacial areas are commensurately small. This should offset the increased pinhole density/probability at reduced film thickness\cite{SchambergerJVSTB12} to maintain a practicable yield of gate-leakage-free devices even in the case where the pinhole density becomes finite but small. This allows for the utilization of films $<$~35~nm as long as the active area is kept small. Atomic-Force Microscopy (AFM) images of our parylene films show an increased surface roughness over thermally grown SiO$_{2}$ however, even films as thin as 10~nm appear conformal with no obvious pinholes on a micrometer scale (see Supplementary Figure~5). Choice of parylene chemistry is also important. We found that the nanoscale film morphology of parylene-C is better than parylene-N (see Supplementary Figure~2), and as a result used parylene-C for all devices discussed from this point onwards.

\begin{figure}
\includegraphics[width=8.5cm]{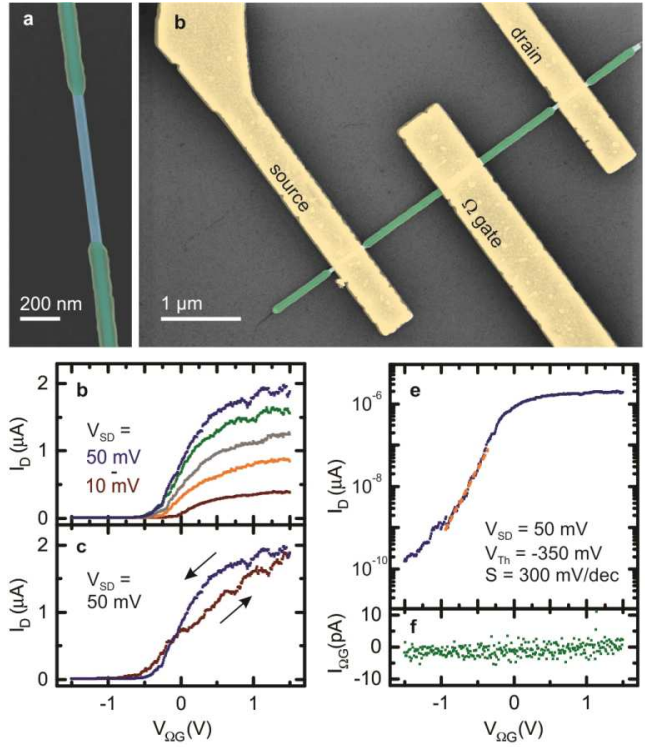}
\caption{(a) False-color SEM image of an InAs nanowire with a $\sim20$~nm parylene-C coating removed in selected areas by oxygen plasma etching. (b) False-color SEM image of a top-gated InAs NWFET with a conformal parylene gate dielectric. InAs is shown in blue, parylene in green and metal in yellow. (c) $I_{D}$ vs $V_{\Omega G}$ for different $V_{SD}$ for the device in (b). (d) The $V_{SD} = 50$~mV data from (c) plotted for both sweep directions of $V_{\Omega G}$ to characterise gate hysteresis. (e) $I_{D}$ vs $V_{\Omega G}$ on a logarithmic scale to obtain the sub-threshold slope $S$ (orange dashed line). (f) Corresponding gate leakage current $I_{G}$ vs $V_{\Omega G}$. All data obtained at room temperature.}
\end{figure}

A thin conformal gate insulator is necessary for more advanced nanowire devices and high-performance gating,\cite{KhanalNL07} e.g., $\Omega$-gates (see Fig.~3),\cite{PfundAPL06} wrap-gates,\cite{StormNL12, BurkeNL15, ThelanderTED08} and gate-all-around (GAA) structures (see Fig.~4). Each of these devices feature a nanowire with a conformal gate insulator layer oriented horizontally on the substrate. An $\Omega$-gate device has a single metal gate strip running across the substrate and over the nanowire,\cite{PfundAPL06} a wrap-gate device has conformal gate metallization deposited on the parylene prior to placement on the device substrate,\cite{StormNL12} and our GAA device has two aligned metal gate strips, one running underneath the nanowire, and one over it, to form the structure shown inset to Fig.~4(a). We avoid calling our GAA structure a `wrap-gate' transistor because of the possibility of small voids under the nanowire due to shadowing during metal evaporation and to distinguish it from the truly conformal wrap-gates of earlier work.\cite{StormNL12, BurkeNL15} In the results that follow we first demonstrate a parylene-insulated $\Omega$-gate device (Fig.~3) and then a parylene-insulated gate-all-around device (Fig.~4). A wrap-gate device similar to that reported by Storm {\it et al.}\cite{StormNL12} should be possible; we opted instead for the GAA device to point out a new approach to making concentrically-gated nanowire transistors. We developed this as a new alternative for instances where sputtering of wrap-gate metal is impractical and/or very short concentric gate structures are required because short gate lengths are challenging with wrap-gate nanowire transistors.\cite{StormNL12, BurkeNL15}

The devices in Figs.~3 and 4 both rely on the ability to deposit a thin conformal parylene layer onto nanowires standing vertically on their growth substrate, transfer them to a separate device substrate, and then etch back the parylene in selected locations to enable source and drain contacts to be made. To demonstrate this capability, Figure~3(a) shows an SEM image of an InAs nanowire with a $\sim20$~nm thick patterned conformal Parylene-C coating featuring a short segment where the parylene film has been etched back using EBL-defined oxygen plasma etching. A nice feature is that insulator patterning does not require a HF-based etchant as for oxides. This also removes the need for the \ce{HfO2} etch-barrier deposited over the \ce{SiO2} back-gate insulator in earlier work.\cite{StormNL12} Figure~3(b) shows an SEM image of a completed device. The conformal parylene coating is etched at the source and drain contacts but not underneath the $\Omega$-gate. All three electrodes are defined in the same EBL and metal deposition process. The electrical performance from a nominally identical device is shown in Figure~3(c-f). The gate reliably switches between clear on- and off-states with a channel resistance of $25$~k$\Omega$ and a threshold voltage $V_{Th}\sim-350$~mV. Measurements of $I_{D}$ versus $V_{SD}$ at fixed $\Omega$-gate voltage $V_{\Omega G}$ (see Supplementary Fig.~3) are linear, indicative of high-quality ohmic contacts to the nanowire and complete removal of the parylene insulator at the contact locations during the patterned oxygen plasma etch step. The $\Omega$-gate shows slight hysteresis; a topic we address in more detail later on. The gate characteristics in Fig.~3(e) demonstrate a room-temperature sub-threshold swing $S = 300$~mV/dec and on-off ratio of $\sim10^4$, competitive with nanowire wrap-gate transistors, where sub-threshold swings typically range from $100$ to $750$~mV/dec.\cite{StormNL12, ThelanderTED08, RehnstedtTED08, FrobergEDL08, RehnstedtEL08, TanakaAPEX10} The same on-off ratio is achieved for SiO$_{2}$-insulated back-gated devices using nanowires from the same growth shown in Supplementary Figure S8. Finally, Fig.~3(f) shows no notable gate leakage current over the full $V_{\Omega G}$ range demonstrating that we can obtain pinhole-free conformal parylene coatings at nanowire scale, i.e., an active gate area of order $10^{-13}$~m$^{2}$.

\begin{figure}
\includegraphics[width=8.5cm]{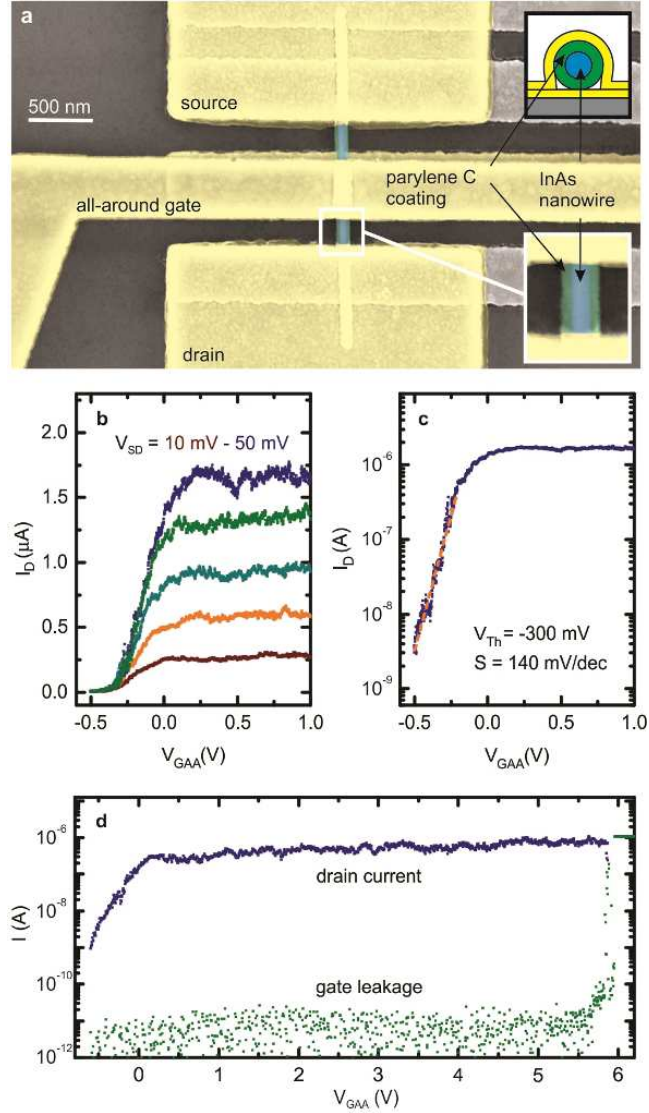}
\caption{(a) False-color SEM image of a gate-all-around (GAA) nanowire transistor with a $\sim20$~nm thick parylene gate dielectric. Schematic cross-section diagram is shown in the inset, highlighting the small voids due to nanowire shadowing during evaporation that make the gate metallization slightly non-conformal. InAs is shown in blue, parylene in green and metal in yellow. (b/c) $I_{D}$ versus gate voltage $V_{GAA}$ for different $V_{SD}$ on a (b) linear and (c) log scale in $I_{D}$. (d) $I_{D}$ (blue/left axis) and gate leakage current $I_{G}$ (green/right axis) vs $V_{GAA}$ from a nominally identical device. All data obtained at room temperature.}
\end{figure}

Figure~4(a) shows an SEM image of our completed gate-all-around (GAA) structure. Fabrication for this device began by defining arrays of fifteen $400$~nm-wide, $30$~nm-thick Ti/Au strips with $200$~nm spacing. Nanowires were precisely placed on top of these strips, perpendicular to the strip orientation. This was achieved by defining a set of $200$~nm-wide trenches in a $300$~nm thick EBL resist, performing random nanowire deposition, and then brushing the substrate parallel to the trench direction to cause some of the nanowires to fall into the trenches.\cite{LimSmall10,LardNL14} The EBL resist was then dissolved, removing any nanowire that has not ended up in a trench in the PMMA and thereby adhered electrostatically to the substrate. EBL was used to pattern a Ti/Au gate strip in Fig.~4(a). This gate strip was aligned to one of the underlying Ti/Au strips to form a gate-all-around structure. In a final EBL step, we plasma-etched the parylene and deposited Ni/Au at the source and drain contacts to complete the device structure. Further details of this fabrication process are given in the Methods.

\begin{figure*}
\includegraphics[width=17cm]{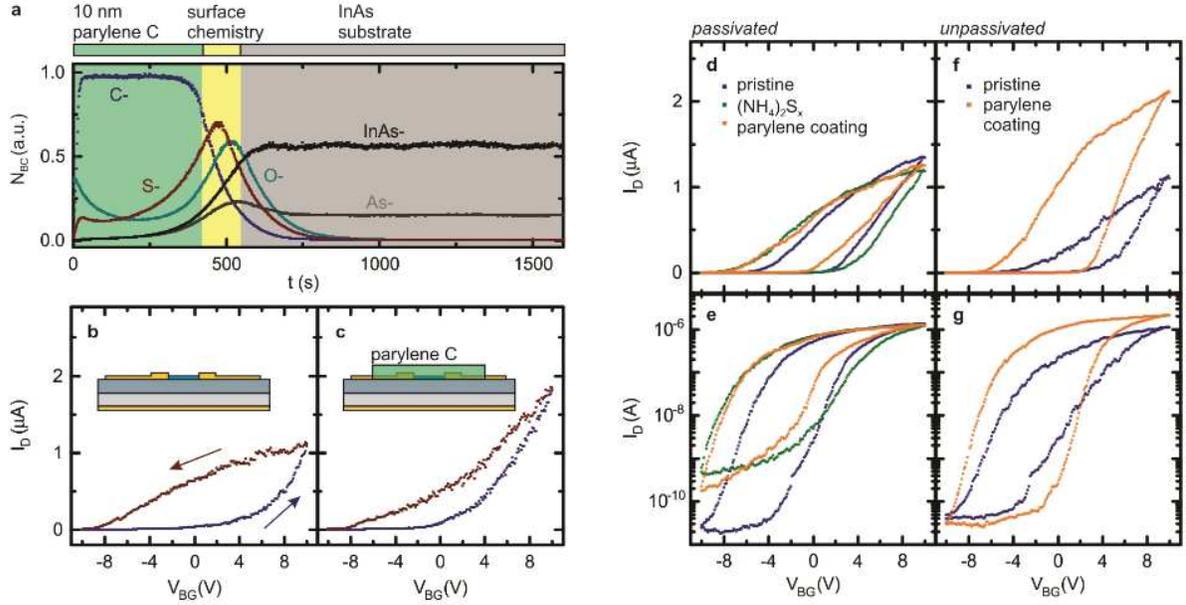}
\caption{(a) Plot of background corrected counts $N_{BC}$ vs sputter time $t$ for time-of-flight secondary ion mass spectrometry (ToF-SIMS) of a $1$~cm$^2$ piece of InAs treated with \ce{(NH4)2S_{x}} immediately prior to deposition of $\sim 10$~nm of parylene-C demonstrating that the passivation treatment survives parylene deposition. (b/c) Plots of $I_D$ vs $V_{BG}$ obtained for a (b) pristine and (c) parylene-C coated back-gated NWFET. Both measurements obtained under room ambient conditions. Sweeps towards more positive $V_{BG}$ presented in red and back towards more negative $V_{BG}$ in blue. (d-g) Plots of $I_D$ vs $V_{BG}$ for passivated and parylene-coated back-gated NWFETs with (d) passivated device linear-$I_D$ scale, (e) passivated device log-$I_D$ scale, (f) unpassivated device linear-$I_D$ scale and (g) unpassivated device log-$I_D$ scale. Data is presented at different points in the process with pristine devices in blue, passivated and uncoated devices in green, and passivated/unpassivated devices with parylene coating in orange. All data obtained under vacuum at room temperature. }
\end{figure*}

Figures~4(b-d) shows the electrical performance of our GAA device. The channel resistance of $30$~k$\Omega$ is only slightly higher than for the $\Omega$-gate device but the gate performance is markedly better with a sub-threshold slope $S = 140$~mV/dec. This is within a factor of $3$ of the room-temperature thermal limit ($60$~mV/dec) and competitive with the best wrap-gate devices, which give $S \sim 100$~mV/dec.\cite{ThelanderEDL08} The gate threshold voltage $V_{Th} = -300$~mV is comparable to the $\Omega$-gate device, indicating no significant differences in surface charge at the parylene interfaces between the two devices. The field-effect mobility $\mu_{FE}$ for the parylene-insulated wrap-gate device was estimated using the cylinder-in-cylinder capacitance\cite{Young_textbook_2004} to be $\mu_{FE}$ = (1600 $\pm$ 300)~cm$^{2}$/Vs. This compares well with the SiO$_{2}$-insulated back gated devices using nanowires from the same growth yielding $\mu_{FE}$ = (1400 $\pm$ 200)~cm$^{2}$/Vs using the cylinder-on-plane capacitance model\cite{Ramo_book_1993} (data shown in Supplementary Figure S8). The obtained value in this work is higher than $\mu_{FE}$ = 109~cm$^{2}$/Vs reported for Al${_2}$O$_{3}$ insulated wrap-gated InAs devices reported by Storm \textit{et al.}\cite{StormNL12}. This may be attributed to the high charge-trap density in the high-$\kappa$ dielectric. As a test for the integrity of the $20$~nm thick parylene insulator we finished the measurement by deliberately pushing the gate voltage to dielectric breakdown as shown in Fig.~4(d). The gate leakage current remains below $10$~pA for $V_{GAA}~<~+5.5$~V and increases drastically at $V_{GAA}\sim+6$~V causing irreversible failure of the parylene insulator. The calculated breakdown field accounting for geometry is $240-300$~mV/nm, which compares well with previous measurements of the breakdown field for Parylene-C thin-films.\cite{HeidCDBE16}

We finish by exploring the potential for using parylene to obtain InAs NWFETs with chemically-treated InAs surfaces. The possible benefits are surface-state passivation for improved gate performance\cite{HangNL08, SunNL12} and threshold voltage tuning using charged adsorbate molecules.\cite{CheungACSNano15, KobayashiNM04} Both are generally precluded for an ALD-deposited oxide because the surface chemistry blocks the ALD process\cite{FarmSST12, DongSciRep14, AvilaAMI14} or the first ALD half-cycle reaction attacks the adsorbed molecules via the interfacial `self-cleaning' effect in ALD.\cite{MilojevicAPL08,HinkleAPL08} In contrast, the physisorption of parylene films is compatible with existing surface chemistry, as shown for alkanethiol monolayers on gold/silver by Vaeth {\it et al.}\cite{VaethLangmuir00} We begin by focussing on the classic \ce{(NH4)2S_{x}} surface treatment commonly used for passivating InAs nanowire contacts,\cite{SuyatinNanotech07} and previously explored for other III-V devices.\cite{FanJJAP88, LiAPL07, CarradJPCM13} We first established that InAs surface treatment remains intact after parylene deposition -- this was untested for III-V surfaces; the only prior literature is for Au/Ag\cite{VaethLangmuir00} where the surface-thiol bonding is stronger. \ce{(NH4)2S_{x}} treatment for III-Vs is well-known to be relatively short lived as a result.\cite{SuyatinNanotech07, CarradJPCM13} Figure~5(a) shows time-of-flight secondary ion mass spectrometry (ToF-SIMS) data for a $1$~cm$^2$ piece of InAs treated with \ce{(NH4)2S_{x}} immediately prior to deposition of $\sim 10$~nm of Parylene-C. Air exposure between passivation and pump-down of the deposition chamber was at most a few minutes. The ToF-SIMS data shows a clear sulfur peak coinciding with a sharp drop in carbon signal and sharp rise in InAs signal after a sputtering time of $\sim500$~s. This is consistent with \ce{(NH4)2S_{x}} treatment remaining intact after parylene deposition. An oxygen signal is also observed at the InAs surface, suggesting some reoxidation of the InAs surface upon air exposure after passivation. The oxygen-to-sulfur ratio is much higher for an equivalent sample without parylene measured as control (see Supplementary Fig.~6). The control sample had longer air exposure as its surface was not protected by a parylene overlayer during transport for analysis. Our InAs nanowires will have wurtzite crystal structure and also present differently indexed facets. However, because parylene physisorbs to the surface, we do not expect radically different results for InAs nanowire over InAs wafer.

The influence of parylene deposition with and without \ce{(NH4)2S_{x}} treatment on the electrical charateristics is shown in Figs.~5(b-g). We began with a control study performed by making a back-gated InAs NWFET, measuring it under ambient conditions (Fig.~5(b)), coating with a parylene-C and then re-measuring it (Fig.~5(c)). The parylene layer reduces the gate hysteresis without adversely affecting the on- or off-current. The reduced hysteresis is likely due to the parylene deposition process first removing surface-adsorbed \ce{H2O} and then encapsulating against \ce{H2O} readsorption. This result is consistent with recent work by Ullah {\it et al.}\cite{UllahNanotech17} where \ce{H2O} was identified as the key atmospheric contributor to gate hysteresis. We performed the \ce{(NH4)2S_{x}} treatment study with two separate sample sets. Figure~5(d/e) shows $I_{D}$ versus $V_{BG}$ for back-gated InAs NWFETs where the device was made, measured (blue trace), treated with \ce{(NH4)2S_{x}}, measured (green trace), coated with parylene-C and measured again (orange trace). These measurements were performed under vacuum to exclude atmospheric variations as a possible cause of changes in hysteresis and limit surface reoxidation. Figure~5(f/g) shows a corresponding control sample where \ce{(NH4)2S_{x}} treatment was not performed. Data obtained for another two treated and two control samples is shown in Supplementary Figure~7 and 8 to demonstrate reproducibility.

Figure~5(c/d) shows that sulfur treatment of the channel leads to a degradation in on-current, increased off-current, and increased hysteresis. Our data is consistent with our earlier work on GaAs,\cite{CarradJPCM13} where \ce{(NH4)2S_{x}} treatment provided little real improvement. Addition of parylene over the \ce{(NH4)2S_{x}} treatment maintains the electrical performance changes attributed to \ce{(NH4)2S_{x}} indicating that the achieved surface chemistry remains intact under the parylene film. The slight increase in on-current and decrease in off-current observed after the parylene deposition may arise from improved gate coupling due to the added dielectric coating.\cite{HeedtNSc15} Consistent with this interpretation, a significant improvement in on-off ratio is observed in a control experiment where pristine nanowires are coated with parylene-C without prior treatment (see Fig.~5(f/g)). The data presented in Fig.~5(d-g) thus confirms that the \ce{(NH4)2S_{x}} treatment on the nanowires is retained through parylene deposition, as it is for the large-area sample in Fig.~5(a). We are not able to confirm this by other means, e.g., ToF-SIMS, due to the tiny interfacial area of our nanowire transistors. Notably, the measurements performed under vacuum show no significant reduction of hysteresis upon the application of the parylene coating. This strengthens the interpretation that the observed improved hysteresis in air is to be attributed to removal and blockage of moisture at the nanowire surface (see Fig~5(b/c)). Under vacuum, we expect the hysteresis to be dominated by charge trapping at the \ce{SiO2}/nanowire interface. Our results with \ce{(NH4)2S_{x}} enable us to address the question recently raised by Holloway {\it et al.}\cite{HollowaySST16} regarding whether \ce{(NH4)2S_{x}} might be a good alternative to 1-octadecanethiol (ODT) for nanowire gate passivation. The data in Fig.~5(d-g) suggests this would not be the case, which is consistent with our earlier work on III-Vs\cite{CarradJPCM13} also. That said, we certainly do find that \ce{(NH4)2S_{x}} treatment does improve the source and drain contacts to InAs nanowires, in agreement with Suyatin {\it et al.}\cite{SuyatinNanotech07}

We also made some attempts with treatments beyond \ce{(NH4)2S_{x}}, including ODT for passivation\cite{SunNL12} and 1H,1H,2H,2H-Perfluorodecanethiol (PFDT) for threshold voltage shifting,\cite{KobayashiNM04} albeit with little success (see Supplementary Information). With ODT in particular, literature suggests that careful elimination of water and oxygen is vital,\cite{SunNL12} and that little benefit is obtained if this is not the case.\cite{HollowaySST16} Further attempts towards devices featuring thiolated self-assembled monolayers under a parylene insulator would entail an oxygen-free passivation capacity well beyond that presently available to us. However, the design of our parylene deposition system would enable this given the necessary atmospheric control infrastructure. Our deposition chamber is small and able to be isolated from the deposition system for removal. Using a glovebox system with a sufficiently large airlock, one could transfer the deposition chamber inside, purge, perform ODT passivation using carefully dehydrated and deoxygenated reagents following the protocol developed by Sun {\it et al.}\cite{SunNL12}, load the sample directly into the deposition chamber inside the glovebox, seal the deposition chamber, remove it from the glovebox and reattach it to the parylene system. One could then pump the adjacent process line spaces, open the valves to evacuate the deposition chamber, and deposit parylene directly onto the passivated surface. Implemented carefully, this could be a $100\%$ \ce{H2O}- and \ce{O2}-free process to obtain chemically-passivated, parylene-insulated gate interfaces for $\Omega$-gate and gate-all-around nanowire transistors.

In summary, we demonstrated the use of parylene as a polymer gate insulator in nanowire field-effect transistors and potentially other nanoscale electronic devices. Parylene can be deposited from the gas-phase at room temperature meaning it is readily substituted for oxide insulators deposited by atomic layer deposition. This could be particularly useful in cases where ALD is unviable, either due to the elevated temperatures or an incompatibility with the chemical bonding process involved in ALD, e.g., to retain surface chemistry. We presented a method for reliable and controlled deposition of parylene films with thicknesses $< 100$~nm onto III-V nanowires both standing vertically on a growth substrate or after transfer to sit horizontally on a device substrate.\cite{Gluschke18} Conformally-coated nanowires can be obtained in the former case. We demonstrated these can be used to produce functional $\Omega$-gate and gate-all-around structures with sub-threshold swings as low as $140$~mV/dec and on/off ratios exceeding $10^3$ at room temperature. The gate-all-around structure was produced using a novel fabrication strategy designed to overcome some of the limitations of lateral wrap-gate nanowire transistors\cite{StormNL12, BurkeNL15} without sacrificing the strong concentric gating. Finally, we explored the deposition of parylene over \ce{(NH4)2S_{x}}-treated nanowire surfaces as a route to nanowire transistors with chemically-passivated semiconductor-insulator interfaces. We used ToF-SIMS to show that \ce{(NH4)2S_{x}} treatment of InAs survives parylene deposition. We find that \ce{(NH4)2S_{x}} treatment of nanowire transistors leads to a reduced on/off ratio with increased hysteresis. Depositing parylene over the surface chemistry largely maintains the changed electrical characteristics arising from \ce{(NH4)2S_{x}} treatment. However, the on/off ratio can be improved by a parylene coating regardless. Our results highlight the potential for parylene as an alternative insulator in nanowire transistors, but also in nanoscale devices more broadly. Parylene's solvent resistance makes it amenable to standard lithographic techniques to obtained patterned etching via oxygen plasma. The ability to combine ultra-thin patterned parylene films into nanoscale electronic devices also offers new applications opportunities in nanobioelectronics. Our demonstration here that parylene, a material with wide-ranging industrial applications already, can be deposited in high-quality ultra-thin films that are then patterned and used for electronic functionality of nanoscale devices not only adds new applications scope to this material but adds a major new material option to the nanoscale device toolbox.

{\bf Methods}\\
{\it Nanowire growth.} The InAs wurtzite nanowires were grown by metal organic vapor phase epitaxy (MOVPE) using Au aerosols with a nominal diameter of $30$~nm as seed particles. The method is similar to that presented elsewhere.\cite{LehmannNL13} An Aixtron $3 \times 2$'' close coupled shower-head system (CCS) was operated at a total carrier gas flow $8$~slm and total reactor pressure $100$~mbar. Growth commenced with a $10$~min anneal in \ce{AsH3}/\ce{H2} ambient at $550^{\circ}$C with \ce{AsH3} molar fraction $\chi_{AsH3}~= ~2.5\times10^{-3}$ before setting the temperature to $470^{\circ}$C for nanowire growth. A short InAs stem was grown at \ce{(CH3)3In} molar fraction of $\chi_{TMIn}~=~1.8\times10^{-6}$ for $180$~s with $\chi_{AsH3}~=~1.2\times10^{-4}$ before the wurtzite InAs nanowire was grown with $\chi_{AsH3}~=~2.3\times10^{-5}$ for $60$~minutes. Growth was terminated by cutting the \ce{(CH3)3In} supply and cooling under \ce{AsH3}/\ce{H2} ambient to $300^{\circ}$C.

{\it Parylene deposition.} $100~\pm~5$~mg of {\it SCS Coatings} DPX-N dimer and {\it Curtiss Wright} C dimer were used to deposit the parylene N and C films. Prior to deposition the CVD system was evacuated to $5\times10^{-4}$~mbar. Valves A and C (as depicted in Fig.~1) remained closed while Valves B, D, and E were kept open. The furnace was heated to $\sim700^{\circ}$C for parylene N and $\sim680^{\circ}$C for parylene C. We waited approximately two hours for all temperatures to stabilize before heating the tubing between the sublimation chamber and furnace to $\sim100^{\circ}$C and the portion between the furnace and deposition chamber to $45^{\circ}$C. This suppresses recrystallization of sublimated dimer prior to reaching the furnace and premature deposition of parylene prior to reaching the deposition chamber. The dimer was heated to $115^{\circ}$C to initiate the deposition. Valve A was opened and Valve B closed to direct the monomer flow at the sample as indicated by the green arrows in Figure~1. The deposition rate $\sim0.5$~\AA/s at $\sim1\times10^{-2}$~mbar was monitored with a {\it Inficon} Q-pod quartz crystal micro-balance. Valve B was opened and A closed to force the gas flow along the path indicated by the orange arrows in Figure~1 when the desired film thickness was reached. The system was subsequently left to cool overnight under vacuum before extracting the sample.

{\it Parylene patterning via plasma etching.} The parylene films were patterned in a {\it Denton Vacuum} PE-$250$ oxygen plasma etcher at $50$~W RF power at a pressure of $340$~mTorr. The typical etch rate was $\sim3$~nm/min for parylene C and $\sim5$~nm/min for parylene N, which is comparable to photoresist under similar conditions.

{\it Device substrates.} The devices were fabricated on highly doped Si substrates with a $\sim140$~nm thick thermally grown \ce{SiO2} layer and $10$~nm of ALD deposited \ce{HfO2}. The \ce{HfO2} exists for HF-etch protection and is an artefact from other projects.~\cite{StormNL12} Alignment and bond pads were defined prior to nanowire deposition by electron-beam and UV photolithography.

{\it Top-gated device (Figure~2).} Nanowires were dry transferred to the device substrate using clean-room tissue and located using darkfield optical microscopy. Source and drain contacts were exposed by EBL ($\sim300$~nm $950$k PMMA A$5$ resist, $20$~kV beam energy, $20~\mu$m aperture). The contact regions were passivated with \ce{(NH4)2S_{x}} solution\cite{SuyatinNanotech07} prior to metallization by thermal evaporation ($6/134$~nm, Ni/Au, lift off in $60^{\circ}$C acetone for $20$~min). A $70$~nm parylene~N layer was deposited as described above. The bond pads were exposed in $\sim1.3~\mu$m thick {\it Shipley} S$1813$ photo resist to remove the underlying parylene layer in an $18$~min oxygen plasma etch ($50$~W, $340$~mTorr). The photoresist was then removed in a $10$~min acetone bath. In a second EBL step, the top-gate was created ($\sim300$~nm $950$k PMMA A$5$ resist, $20$~kV beam energy, $20~\mu$m aperture, deposit $6/134$~nm Ni/Au, lift off in $60^{\circ}$C acetone for $20$~min).

{\it $\Omega$-gated device (Figure~3).} The nanowires were conformally coated with $20-25$~nm of parylene~C whilst standing on the growth substrate. The nanowire substrate was rotated at $\sim10$~rpm and tilted at a $30^{\circ}$ angle to the process line axis of the parylene deposition system. The coated nanowires were then dry transferred to a device substrate. To selectively remove the parylene, the device was coated with $\sim300$~nm of $950$k PMMA A$5$ EBL resist,
the source and drain regions were exposed by EBL, and then the sample was oxygen plasma etched ($50$~W, $340$~mTorr) for $8$~min. The resist was removed in a $20$~min acetone bath, prior to spinning a fresh layer of EBL resist to expose the source, drain and gate electrodes. The sample was treated in \ce{(NH4)2S_{x}} solution\cite{SuyatinNanotech07} prior to metal deposition ($6/134$~nm, Ni/Au, lift off in $60^{\circ}$C acetone for $20$~min).

{\it Gate-all-around device (Figure~4).} Arrays of $400$~nm wide bottom gates were defined by EBL and thermally evaporated ($5$~nm Ti, $30$~nm Au) prior to nanowire deposition. The sample was then coated with $\sim300$~nm thick EBL resist and arrays of $200$~nm wide trenches\cite{LardNL14} oriented perpendicular to the bottom gates were exposed. The nanowires were transferred from the parylene-coated growth substrate onto the patterned resist using clean-room tissue. The sample surface was wet with a drop of $2$-propanol before brushing the nanowires into the trenches using clean-room tissue. The EBL resist was removed and an upper gate-strip parallel to one of the bottom gate-strips underneath the nanowire was created in a third EBL step ($6$~nm Ni, $134$~nm Au). The parylene coating of the nanowire was selectively removed in the contact area using oxygen plasma etching (RF power $50$~W, $340$~mTorr, $8$~min) after the contacts were exposed by EBL. The resist is removed and a fresh resist layer is spun before creating the contacts to the nanowire in a final EBL step because the plasma etching the attacks the EBL resist as well as the parylene. The sample was treated in \ce{(NH4)2S_{x}} solution\cite{SuyatinNanotech07} prior to metal deposition ($6/134$~nm, Ni/Au, lift off in $60^{\circ}$C acetone for $20$~min).

{\it Electrical measurements.} All measurements were conducted at room temperature and under vacuum unless stated otherwise. A {\it Keithley} $2401$ SourceMeter was used to apply a dc-voltage $V_{SD}$ between $10$ and $50$~mV to the nanowire source. The resulting current $I_{D}$ was measured using a {\it Keithley} $6517$A Electrometer at the nanowire drain. The gate voltage $V_{BG}$, $V_{TG}$, $V_{\Omega}$ or $V_{GAA}$ was applied using a \textit{Keithley} $2400$ SourceMeter, which enabled continuous monitoring of the gate leakage current $I_{G}$.

{\it ToF-SIMS measurements.} The ToF-SIMS measurements were performed using an \textit{ION-TOF} TOF-SIMS$^{5}$. A 300$\times$ 300 $\mu$m$^{2}$ area of the sample was sputtered with 250~eV, 13.4~nA Cs$^{+}$ ions. A 100$\times$ 100 $\mu$m$^{2}$ region was investigated with a primary beam of 15~keV, 0.4~pA Bi$_{3}^{+}$ ions in negative polarity.

{\bf Supporting Information.} Additional fabrication details, scanning electron microscopy images, atomic force microscopy images, capacitance measurements, electrical and ToF-SIMS data, as well as a discussion of attempts to apply ODT passivation. This material is available free of charge via the Internet at http://pubs.acs.org.

\acknowledgement

This work was funded by the Australian Research Council (ARC) under DP170102552 and DP170104024, UNSW Goldstar Scheme, NanoLund at Lund University, Swedish Research Council, Swedish Energy Agency (Grant No. 38331-1) and Knut and Alice Wallenberg Foundation (KAW). APM acknowledges an ARC Future Fellowship (FT0990285). This work was performed in part using the NSW node of the Australian National Fabrication Facility (ANFF). We thank Felix Richter for assistance with parts of the surface chemistry studies and Bill Gong from the Mark Wainwright Analytical Centre at the University of New South Wales for performing the ToF-SIMS measurements.


\begin{thebibliography}:

\bibitem{WongMicroelecEng06} Wong, H. and Iwai, H. On the scaling issues and high-$\kappa$ replacement of ultrathin gate dielectrics for nanoscale MOS transistors. {\it Microelec. Eng.} {\bf 2006}, {\it 83}, 1867-1904.

\bibitem{RobertsonMSER15} Robertson, J. and Wallace, R.M. High-K materials and metal gates for CMOS applications. {\it Mat. Sci. Eng. R} {\bf 2015}, {\it 88}, 1-41.

\bibitem{RielMRS14} Riel, H.; Wernersson, L.-E.; Hong, M.; del Alamo, J.A. III-V compound semiconductor transistors -- from planar to nanowire structures. {\it MRS Bulletin} {\bf 2014}, {\it 39}, 668-677.

\bibitem{DayehAPL07} Dayeh, S.A.; Soci, C.; Yu, P.K.L.; Yu, E.T.; Wang, D. Influence of surface states on the extraction of transport parameters from InAs nanowire field effect transistors. {\it Appl. Phys. Lett.} {\bf 2007}, {\it 90}, 162112.

\bibitem{RoddaroAPL08} Roddaro, S.; Nilsson, K.; Astromskas, G.; Samuelson, L.; Wernersson, L.-E.; Karlstr\"{o}m, O.; Wacker, A. InAs nanowire metal-oxide-semiconductor capacitors. {\it Appl. Phys. Lett.} {\bf 2008}, {\it 92}, 253509.

\bibitem{HollowayJAP13} Holloway, G.W.; Song, Y.; Haapamaki, C.M.; LaPierre, R.R.; Baugh, J. Trapped charge dynamics in InAs nanowires. {\it J. Appl. Phys.} {\bf 2013}, {\it 113}, 024511.

\bibitem{LiuAPL11} Liu, H.; Ye, P.D. Atomic-layer-deposited \ce{Al2O3} on \ce{Bi2Te3} for topological insulator field-effect transistors. {\it Appl. Phys. Lett.} {\bf 2011}, {\it 99}, 052108.

\bibitem{DayehJVSTB07} Dayeh, S.A.; Soci, C.; Yu, P.K.L.; Yu, E.T.; Wang, D. Transport properties of InAs nanowire field effect transistors: The effects of surface states. {\it J. Vac. Sci. Technol. B} {\bf 2007}, {\it 25}, 1432–1436.

\bibitem{FleischliLangmuir10} Fleischli, F.D.; Su\`{a}rez, S.; Schaer, M.; Zuppiroli, L. Organic thin-film transistors: The passivation of the dielectric-pentacene interface by dipolar self-assembled monolayers. {\it Langmuir} {\bf 2010}, {\it 26}, 15044-15049.

\bibitem{Illarionov2DM17} Illarionov, Y.Y.; Knobloch, T.; Waltl, M.; Rzepa, G.; Pospischil, A.; Polyushkin, D.K.; Furchi, M.M.; Mueller, T.; Grasser, T. Energetic mapping of oxide traps in \ce{MoS2} field-effect transistors. {\it 2D Mater.} {\bf 2017}, {\it 4}, 025108.

\bibitem{BonmannJVSTB17} Bonmann, M.; Vorobiev, A.; Stake, J.; Engstr\"{o}m, O. Effect of oxide traps on channel transport characteristics in graphene field effect transistors. {\it J. Vac. Sci. Technol. B}, {\bf 2017}, {\it 35}, 01A115.

\bibitem{MilojevicAPL08} Milojevic, M.; Hinkle, C.L.; Aguirre-Tostado, F.S.; Kim, H.C.; Vogel, E.M.; Kim, J.; Wallace, R.M. Half-cycle atomic layer deposition reaction studies of \ce{Al2O3} on \ce{(NH4)2S} passivated GaAs(100) surfaces. {\it Appl. Phys. Lett.} {\bf 2008}, {\it 93}, 252905.

\bibitem{HinkleAPL08} Hinkle, C.L.; Sonnet, A.M.; Vogel, E.M.; McDonnell, S.; Hughes, G.J.; Milojevic, M.; Lee, B.; Aguirre-Tostado, F.S.; Choi, K.J.; Kim, H.C.; Kim, J.; Wallace, R.M. GaAs interfacial self-cleaning by atomic layer deposition. {\it Appl. Phys. Lett.} {\bf 2008}, {\it 92}, 071901.

\bibitem{FarmSST12}	F\"{a}rm, E.; Vehkam\"{a}ki, M.; Ritala, M.; Leskel\"{a}, M. Passivation of copper surfaces for selective-area ALD using a thiol self-assembled monolayer. {\it Semicond. Sci. Technol.} {\bf 2012}, {\it 27}, 074004.

\bibitem{DongSciRep14} Dong, W.; Zhang, K.; Zhang, Y.; Wei, T.; Sun, Y.; Chen, X.; Dai, N. Application of three-dimensionally area-selective atomic layer deposition for selectively coating the vertical surfaces of standing nanopillars. {\it Sci. Rep.} {\bf 2014}, {\it 4}, 4458.

\bibitem{AvilaAMI14} Avila, J.R.; DeMarco, E.J.; Emery, J.D.; Farha, O.K.; Pellin, M.J.; Hupp, J.T.; Martinson, A.B.F. Real-time observation of atomic layer deposition inhibition: Metal oxide growth on self-assembled alkanethiols. {\it ACS Appl. Mater. Interfaces} {\bf 2014}, {\it 6}, 11891–11898.

\bibitem{FortinCM02} Fortin, J.B.; Lu, T.-M. A model for the chemical vapor deposition of poly(para-xylylene) (parylene) thin films. {\it Chem. Mater.} {\bf 2002}, {\it 14}, 1945-1949.

\bibitem{VaethLangmuir00} Vaeth K.M.; Jackman, R.J.; Black, A.J.; Whitesides, G.M.; Jensen, K.F. Use of microcontact printing for generating selectively grown films of poly(p-phenylene vinylene) and parylenes prepared by chemical vapour deposition. {\it Langmuir} {\bf 2000}, {\it 16}, 8495-8500.

\bibitem{ChangLangmuir07} Chang, T.Y.; Yadav, V.G.; De Leo, S.; Mohedas, A.; Rajalingam, B.; Chen, C.-L.; Selvarasah, S.; Dokmeci, M.R.; and Khademhosseini, A. Cell and protein compatibility of parylene-C surfaces. {\it Langmuir} {\bf 2007}, {\it 23}, 11718-11725.

\bibitem{KhodagholyNatComm13} Khodagholy, D.; Doublet, T.; Quilichini, P.; Gurfinkel, M.; Leleux, P.; Ghestem, A.; Ismailova, E.; Herv\'{e}, T.; Sanaur, S.; Bernard, C.; Malliaras, G.G. {\it In vivo} recordings of brain activigty using organic transistors. {\it Nature Communications} {\bf 2013}, {\it 4}, 1575.

\bibitem{KoitmaeAMI16} Koitm\"{a}e, A.; Harberts, J.; Loers, G.; M\"{u}ller, M.; Bausch, C.S.; Sonnenberg, D.; Heyn, C.; Zierold, R.; Hansen, W.; Blick, R.H. Approaching integrated hybrid neural circuits: Axon guiding on optically active semiconductor microtube arrays. {\it Adv. Mater. Interfaces} {\bf 2016}, {\it 3}, 1600746.

\bibitem{NoyAdvMat10} Noy, A. Bionanoelectronics. {\it Adv. Mater.} {\bf 2011}, {\it 23}, 807-820.

\bibitem{ZhangChemRev16} Zhang, A.; Lieber, C.M. Nano-Bioelectronics. {\it Chem. Rev.} {\bf 2016}, {\it 116}, 215-257.

\bibitem{PodzorovAPL03} Podzorov, V.; Pudalov, V.M.; Gershenson, M.E. Field-effect transistors on rubrene single crystals with parylene gate insulator. {\it Appl. Phys. Lett.} {\bf 2003}, {\it 82}, 1739-1741.

\bibitem{PodzorovAPL03b} Podzorov, V.; Sysoev, S.E.; Loginova, E.; Pudalov, V.M. and Gershenson, M.E. Single-crystal organic field effect transistors with the hole mobility $\sim 8$~cm$^{2}$/Vs. {\it Appl. Phys. Lett.} {\bf 2003}, {\it 83}, 3504-3506.

\bibitem{SabriAPL09} Sabri, S.S.; L\'{e}vesque, P.L.; Aguirre, C.M.; Guillemette, J.; Martel, R.; Szkopek, T. Graphene field effect transistors with parylene gate dielectric. {\it Appl. Phys. Lett.} {\bf 2009}, {\it 95}, 242104.

\bibitem{ChamlagainACSNano14} Chamlagain, B.; Li, Q.; Ghimire, N.J.; Chuang, H.-J.; Perera M.M.; Tu, H.; Xu, Y.; Pan, M.; Xaio, D.; Yan, J.; Mandrus, D.; Zhou, Z. Mobility improvement and temperature dependence in \ce{MoSe2} field-effect transistors on parylene-C substrate. {\it ACS Nano} {\bf 2014}, {\it 8}, 5079-5088.

\bibitem{StormNL12} Storm, K.; Nylund, G.; Samuelson, L.; Micolich, A.P. Realizing lateral wrap-gated nanowire FETs: Controlling gate length with chemistry rather than lithography. {\it Nano Lett.} {\bf 2012}, {\it 12}, 1–6.

\bibitem{RappTSF12} Rapp, B.E.; Voigt, A.; Dirschka, M.; L\"{a}nge, K. Deposition of ultrathin parylene C films in the range of $18$~nm to $142$~nm: Controlling the layer thickness and assessing the closeness of the deposited films. {\it Thin Solid Films} {\bf 2012}, {\it 520}, 4884-4888.

\bibitem{FortinBook03} Fortin, J.B.; Lu, T.-M. {\it Chemical Vapor Deposition Polymerization: The Growth and Properties of Parylene Thin Films}, (Springer, New York, 2003).

\bibitem{Gluschke18} Gluschke, J. G.; Richter, F.; Cochrane, J.; Micolich, A. P. Chemical Vapor Deposition of Ultra-Thin Parylene Films for Nanostructures. {\it to be submitted to Rev. Sci. Instrum.} {\bf 2018}.

\bibitem{DuanNat01} Duan, X.; Huang, Y.; Cui, Y.; Wang, J.; Lieber, C.M. Indium phosphide nanowires as building blocks for nanoscale electronic and optoelectronic devices. {\it Nature} {\bf 2001}, {\it 409}, 66-69.

\bibitem{MengJMM08} Meng, E.; Li, P.-Y.; Tai, Y.-C. Plasma removal of parylene C. {\it J. Micromech. Microeng.} {\bf 2008}, {\it 18}, 045004.

\bibitem{SpivackJECS69} Spivack, M.A.; Ferrante, G. Determination of the water vapor permeability and continuity of ultrathin parylene membranes. {\it J. Electrochem. Soc.} {\bf 1969}, {\it 116}, 1592-1594.

\bibitem{SchambergerJVSTB12} Schamberger, F.; Ziegler, A.; Franz, G. Influence of film thickness and deposition rate on surface quality of polyparylene coatings. {\it J. Vac. Sci. Technol. B} {\bf 2012}, {\it 30}, 051801.

\bibitem{KhanalNL07} Khanal, D.R.; Wu, J. Gate coupling and charge distribution in nanowire field effect transistors. {\it Nano Lett.} {\bf 2007}, {\it 7}, 2778-2783.

\bibitem{PfundAPL06} Pfund, A.; Shorubalko, I.; Leturcq, R.; Ensslin, K. Top-gate defined double quantum dots in InAs nanowires. {\it Appl. Phys. Lett.} {\bf 2006}, {\it 89}, 252106.

\bibitem{ThelanderTED08} Thelander, C.; Rehnstedt, C.; Fr\"{o}berg, L.E.; Lind, E.; M{\aa}rtensson, T.; Caroff, P.; L\"{o}wgren, T.; Ohlsson, B.J.; Samuelson, L.; Wernersson, L.E. Development of a vertical wrap-gated InAs FET. {\it IEEE Trans. Electron Devices} {\bf 2008}, {\it 55}, 3030–3036.

\bibitem{RehnstedtTED08} Rehnstedt, C.; M{\aa}rtensson, T.; Thelander, C.; Samuelson, L.; Wernersson, L.-E. Vertical InAs nanowire wrap gate transistors on Si substrates. {\it IEEE Trans. Electron Devices} {\bf 2008}, {\it 55}, 3037-3041.

\bibitem{FrobergEDL08} Fr\"{o}berg, L.E.; Rehnstedt, C.; Thelander, C.; Lind, E.; Wernersson, L.-E.; Samuelson, L. Heterostructure barriers in wrap-gated nanowire FETs. {\it IEEE Electron Device Lett.} {\bf 2008}, {\it 29}, 981-983.

\bibitem{RehnstedtEL08} Rehnstedt, C.; Thelander, C.; Fr\"{o}berg, L.E.; Ohlsson, B.J.; Samuelson, L.; Wernersson, L.-E. Drive current and threshold voltage control in vertical InAs wrap-gate transistors. {\it Electronics Lett.} {\bf 2008}, {\it 44}, 236-237.

\bibitem{TanakaAPEX10} Tanaka, T.; Tomioka, K.; Hara, S.; Motohisa, J.; Sano, E.; Fukui, T. Vertical surrounding gate transistors using single InAs nanowire grown on Si substrates. {\it Appl. Phys. Express} {\bf 2010}, {\it 3}, 025003.

\bibitem{BurkeNL15} Burke, A.M.; Carrad, D.J.; Gluschke, J.G.; Storm, K.; Fahlvik Svensson, S.; Linke, H.; Samuelson, L.; Micolich, A.P. InAs nanowire transistors with multiple, independent wrap-gate segments. {\it Nano Lett.} {\bf 2015}, {\it 15}, 2836-2843.

\bibitem{LimSmall10} Lim, J.K.; Lee, B.Y.; Pedano, M.L.; Senesi, A.J.; Jang, J.-W.; Shim, W.; Hong, S.; Mirkin, C.A. Alignment strategies for the assembly of nanowires with submicron diameters. {\it Small} {\bf 2010}, {\it 6}, 1736-1740.

\bibitem{LardNL14} Lard, M.; ten Siethoff, L.; Generosi, J.; M{\aa}nsson, A.; Linke, H. Molecular motor transport through hollow nanowires. {\it Nano Lett.} {\bf 2014}, {\it 14}, 3041-3046.

\bibitem{ThelanderEDL08} Thelander, C.; Fr\"{o}berg, L.E.; Rehnstedt, C.; Samuelson, L.; Wernersson, L.-E. Vertical enhancement-mode InAs nanowire field-effect transistor with $50$-nm wrap gate. {\it IEEE Electron Device Lett.} {\bf 2008}, {\it 29}, 206–208.

\bibitem{Young_textbook_2004} Young, H.; Freedman, R. 24.4 A Cylindrical Capacitor. In University Physics; Pearson Education: San Francisco, 2004; p 913.

\bibitem{Ramo_book_1993} Ramo, S.; Whinnery, J. R.; van Duzer, T. Fields and Waves in Communication Electronics, 3rd ed.; Wiley: New York, 1993.

\bibitem{HeidCDBE16} Heid, A.; von Metzen, R.; Stett, A.; Bucher, V. Examination of dielectric strength of thin parylene C films under various conditions. {\it Curr. Dir. Biomed. Eng.} {\bf 2016}, {\it 2}, 39-41.

\bibitem{LehmannNL13} Lehmann, S.; Wallentin, J.; Jacobsson, D.; Deppert, K.; Dick, K.A. A general approach for sharp crystal phase switching in InAs, GaAs, InP, and GaP nanowires using only group V flow. {\it Nano Lett.} {\bf 2013}, {\it 13}, 4099–4105.

\bibitem{HangNL08} Hang, Q.; Wang, F.; Carpenter, P.D.; Zemlyanov, D.; Zakharov, D.; Stach, E.A.; Buhro, W.E.; Janes, D.B. Role of molecular surface passivation in electrical transport properties of InAs nanowires. {\it Nano Lett.} {\bf 2008}, {\it 8}, 49–55.

\bibitem{SunNL12} Sun, M.H.; Joyce, H.J.; Gao, Q.; Tan, H.H.; Jagadish, C.; Ning, C.Z. Removal of surface states and recovery of band-edge emission in InAs nanowires through surface passivation. {\it Nano Lett.} {\bf 2012}, {\it 12}, 3378–3384

\bibitem{CheungACSNano15} Cheung, H.-Y.; Yip, S.; Han, N.; Dong, G.; Fang, M.; Yang, Z.-X.; Wang, F.; Lin, H.; Wong, C.-Y.; Ho, J.C. Modulating electrical properties of InAs nanowires {\it via} molecular monolayers. {\it ACS Nano} {\bf 2015}, {\it 9}, 7545-7552.

\bibitem{SuyatinNanotech07} Suyatin, D.B.; Thelander, C.; Bj\"{o}rk, M.T.; Maximov, I.; Samuelson, L. Sulfur passivation for ohmic contact formation to InAs nanowires. {\it Nanotechnology} {\bf 2007}, {\it 18}, 105307.

\bibitem{FanJJAP88} Fan, J.-F.; Oigawa, H.; Nannichi, Y. The effect of \ce{(NH4)2S} treatment on the interface characteristics of GaAs MIS structures. {\it Jap. J. Appl. Phys.} {\bf 1988}, {\it 27}, L1331-L1333.

\bibitem{LiAPL07} Li, J.V.; Chuang, S.L.; Aifer, E.; Jackson, E.M. Surface recombination velocity reduction in type-II InAs/GaSb superlattice photodiodes due to ammonium sulfide passivation. {\it Appl. Phys. Lett.} {\bf 2007}, {\it 90}, 223503.

\bibitem{CarradJPCM13} Carrad, D.J.; Burke, A.M.; Reece, P.J.; Lyttleton, R.J.; Waddington, D.E.J.; Rai, A.; Reuter, D.; Wieck, A.D.; Micolich, A.P. The effect of \ce{(NH4)2Sx} passivation on the (311)A GaAs surface and its use in AlGaAs/GaAs heterostructure devices. {\it J. Phys.: Condens. Matter} {\bf 2013}, {\it 25}, 325304.

\bibitem{KobayashiNM04} Kobayashi, S.; Nishikawa, T.; Takenobu, T.; Mori, S.; Shimoda, T.; Mitani, T.; Shimotani, H.; Yoshimoto, N.; Ogawa, S.; Iwasa, Y. Control of carrier density by self-assembled monolayers in organic field-effect transistors. {\it Nature Materials} {\bf 2004}, {\it 3}, 317-322

\bibitem{UllahNanotech17} Ullah, A.R.; Joyce, H.J.; Tan, H.H.; Jagadish, C.; Micolich, A.P. The influence of atmosphere on the performance of pure-phase WZ and ZB InAs nanowire transistors. {\it Nanotechnology}, {\bf 2017}, {\it 28}, 454001.

\bibitem{ChenNL17} Chen, I.-J.; Lehmann, S.; Nilsson, M.; Kivisaari, P.; Linke, H.; Dick, K.A.; Thelander, C. Conduction band offset and polarization effects in InAs nanowire polytype junctions. {\it Nano Lett.} {\bf 2017}, {\it 17}, 902-908.

\bibitem{DeJarldNL11} DeJarld, M.; Shin, J.C.; Chern, W.; Chanda, D.; Balasundaram, K.; Rogers, J.A.; Li, X. Formation of high aspect ratio GaAs nanostructures with metal-assisted chemical etching. {\it Nano Lett.} {\bf 2011}, {\it 11}, 5259-5263.

\bibitem{Ullah18} Ullah, A.R.; Meyer, F.; Gluschke, J.G.; Naureen, S.; Caroff, P.; Krogstrup, P.; Nyg{\aa}rd, J.; Micolich, A.P. $p$-GaAs nanowire MESFETs with near-thermal limit gating. Manuscript in preparation.

\bibitem{HollowaySST16} Holloway, G.W.; Haapamaki, C.M.; Kuyanov, P.; LaPierre, R.R.; Baugh, J. Electrical characterization of chemical and dielectric passivation of InAs nanowires. {\it Semicond. Sci. Technol.} {\bf 2016}, {\it 31}, 114004.

\bibitem{HeedtNSc15} Heedt, S.; Otto, I.; Sladek, K.; Hardtdegen, H.; Schubert, J.; Demarina, N.; L\"{u}th, H.; Gr\"{u}tzmacher, D.; Sch\"{a}pers, T. Resolving ambiguities in nanowire field-effect transistor characterization. {\it Nanoscale} {\bf 2015}, {\it 7}, 18188-18197.

\end{thebibliography}
\end{document}